\documentclass[prl,twocolumn,showpacs]{revtex4-1}

\usepackage[dvips]{graphicx}
\usepackage[activeacute,english]{babel}
\usepackage{amsmath}
\newcommand{\be}{\begin{equation}}
\newcommand{\ee}{\end{equation}}
\newcommand{\px}{p_{\!_X\!}}
\newcommand{\py}{p_{\!_Y\!}}
\newcommand{\pz}{p_{\!_Z\!}}
\newcommand{\lx}{\Lambda_{\!_X\!}}
\newcommand{\mx}{m_{\!_X\!}}

\begin{document}

\title{Entropy of continuous mixtures and the measure problem
}

\author{Pablo Maynar}
\affiliation{F\'{\i}sica Te\'{o}rica, Universidad de Sevilla,
Apartado de Correos 1065, E-41080, Sevilla, Spain}
\author{Emmanuel Trizac}
\affiliation{LPTMS (CNRS UMR 8626), Universit\'e Paris-Sud, Orsay Cedex, 
F-91405, France}

\begin{abstract}
In its continuous version, the entropy functional measuring the
information content of a given probability density may be plagued by
a ``measure'' problem that results from improper weighting
of phase space. This issue is addressed considering a generic 
collision process whereby a large number of particles/agents
randomly and repeatedly 
interact in pairs, with prescribed conservation law(s). We find 
a sufficient condition
under which the stationary single particle distribution function 
maximizes an entropy-like functional, that is
free of the measure problem. 
This condition amounts to a factorization property of the Jacobian
associated to the binary collision law, from which the proper 
weighting of phase space directly follows.
\end{abstract}

\maketitle

In information theory, the definition of the entropy of a continuous
probability distribution depends on the identification of a relevant
prior, or weighting function \cite{Jaynes,Balian}, that
can prove elusive. To illustrate this point and motivate our
approach, let us
consider an ensemble of particles (each indexed by integer $i$) that can 
exchange some positive quantity $x$ so that $\sum_i x_i$ is fixed: 
two particles $i$ and $j$ chosen at random interact so that 
$x_i \to x_i + \eta$ and $x_j \to x_j -\eta$, provided both quantities remain 
positive. Here, $\eta$
is a fixed small increment, or can be drawn from a prescribed
distribution. Such a model has appeared in different settings:
In the context of mass transport models,
$x$ stands for the mass of the particle \cite{M09}; 
it can also be the position of a composite object in exclusion processes
\cite{Ligget},
the volume 
of some colloidal aggregate \cite{Z99}, the size of a self-assembled polymer
\cite{CS02}, the wealth of an agent 
in a simplistic econophysics framework
\cite{D01}, or an auxiliary quantity used for 
algorithmic purposes, in particular the generation of pseudo-random numbers 
\cite{Criado}. 
Upon iterating the previous ``collision'' rule, it can be shown
that the $x$-distribution reaches the simple stationary probability
density function $\px(x) = \exp(-x)$, fixing for convenience
the mean $x$ to unity \cite{M09}. Following the early work of C. Shannon \cite{Shannon}, 
this result {\em seems} to be readily 
recovered by maximizing the information measure  --or differential entropy--
of the distribution
\be
S_{\text{Shannon}} \,=\,- \int \px(x) \log[\px(x)]\, dx
\label{eq:SShannon}
\ee
under the constraint that $\int \px \, dx=\int x \,\px dx=1$ \cite{rquesha}. 
On the other hand, it is clear that the
process could be equally well described by another quantity $y$ 
(say, the radius of a colloid instead of its volume), with a corresponding
probability density $\py$ such that $\px(x) dx = \py(y) dy$.
However, the formulation (\ref{eq:SShannon}) is not invariant under change
of variable, so that a different and inconsistent distribution would be
found by maximizing $-\int \py \log \py$, even after taking into account
for constraints appropriately. We will refer to 
this latent deficiency, already noted in \cite{Shannon},  
as the ``measure problem''. 
In addition, although the pathological nature of Eq. 
(\ref{eq:SShannon}) is made evident by a change of variable,
it can also be inferred from its dimensional inconsistency.
We conclude that recovering 
the correct (in our example exponential) distribution from maximizing 
(\ref{eq:SShannon}) is coincidental, and that Eq. (\ref{eq:SShannon})
does not provide an admissible information measure. 
We emphasize that Shannon faced the measure problem \cite{Shannon:pb}, 
and concluded that the entropy 
of a continuous distribution is not an absolute measure, but
is relative to the coordinate
system. 
Such a point of view is not acceptable: the entropy should not have
an absolute status for discrete probabilities, and a relative
one for continuous cases.

The mechanism for recovering an absolute 
information measure that is unaffected by a 
parameter change, is clear when the continuous limit is carefully taken
from the situation described by a discrete probability set $\{p_\alpha\}$, 
where the entropy reads $-\sum_\alpha p_\alpha \log p_\alpha$ 
\cite{Shannon,Balian,CT}. In doing so, it is necessary to introduce
the density of points $\mx(x)$ and one obtains \cite{Jaynes}
\be
S \,=\, -\int \px(x) \log\left [\lx(x)\,\px(x) \right]\, dx .
\label{eq:S}
\ee
In the above expression, that seems to have been first derived
and commented by Jaynes \cite{Jaynes,rque2}, 
the quantity $\lx$ can be viewed as 
a weighting function,
and  it is related to the density by $\lx(x) = 1/\mx(x)$. This 
$x$-dependent function
indicates how the space of dynamical variables is resolved \cite{Balian}: 
The larger the density $\mx$, the better the resolution,
which corresponds to a smaller $\lx$.
Since the densities $m$ transform 
under change of variable as the probability densities $p$ do, the 
coordinate dependence of $\Lambda$ cures the measure problem.
It is therefore essential to understand what this dependence is,
a problem that is quite often overlooked in the literature 
\cite{Z99,CS02,overlooked} and that
Jaynes --somewhat ironically--
ascribes to the fact that ``one could not think of anything else
to do'' \cite{Jaynes2}. Consequently, if the density $\mx$ can
be extracted from our knowledge of $x$-space sampling, the measure
problem is solved. This case is that of a ``quenched'' $x$-distribution.
There are nevertheless situations where this knowledge is not
{\it a priori} available, but is encoded in the dynamics of the 
system (``annealed'' $x$-distribution), so that $\mx$ is selected by the
underlying dynamical rules. Our goal here is to understand
that connection in an annealed context, 
in order to set up a clear prescription 
for writing the relevant entropy.

We are now in a position to state the problem more precisely.
We are interested in a population of a large number $N$ of 
particles where a given property $x_i$ (mass, velocity, 
length, color, income etc) 
is attached to each particle $i$. These particles undergo 
repeatedly binary
``collisions'' where pairs selected at random interact such that
$(x_i,x_j) \to (x'_i,x'_j)$. An 
important point is that we assume the existence
of a conservation law  
\be
{\cal C}(x_i)+{\cal C}(x_j) = {\cal C}(x'_i)+{\cal C}(x'_j),
\label{eq:cons}
\ee
where $\cal C$ is a given function.
We shall leave ergodicity issues aside, and consider that the 
functions $x'_i(x_i,x_j),x'_j(x_i,x_j)$, that are not specified, 
are sufficiently mixing
to ensure that all accessible phase space is sampled 
(in general, non uniformly). 
The objective is to answer the following question. 
${\cal Q}$: Can we maximize a functional of the 
form (\ref{eq:S}), under the appropriate constraints that 
$\int \px(x) dx$ and $\int {\cal C}(x) \px(x) dx$ are fixed,
to obtain the steady state probability distribution 
$\px^{st}(x)$, if it exists? If so, simple calculus 
shows that the latter distribution
is of the form
\be
\px^{st}(x) = \alpha \,\lx^{-1}(x) \, \exp[-{\beta \, {\cal C}(x)}]
\label{eq:pstar}
\ee
where $\alpha$ and $\beta$ are irrelevant Lagrange multipliers.
The ensuing problem is then to understand what specifies the 
weighting function $\lx$. Indeed, knowing that ${\cal Q}$
can be answered affirmatively is of little interest
if one does not know the corresponding weighting function 
$\lx(x)$. 

The collision law considered may violate
detailed balance,
and it may involve an additional stochastic parameter $\eta$,
as for instance in the simple example introduced in \cite{Criado},
that we mention as a warm-up exercise:
\be
\Bigl|\begin{array}{c}
x'_1\\x'_2
\end{array}
 \,=\, \Bigl|\, 
\begin{array}{c}
\eta \,(x_1+x_2)/\sqrt{2}\\
\eta \,(x_1-x_2)/\sqrt{2}
\end{array},
\label{eq:criado}
\ee
where $\eta$ equiprobably takes values $\pm 1$.
Likewise, randomness is necessarily introduced for colliding hard bodies,
as a remnant of the impact parameter in a description that only considers
the velocity degrees of freedom, as routinely done in some Monte Carlo
simulation techniques \cite{Bird}. 
It should be clear from the outset that the conserved quantity is in general
not exponentially distributed, as our simplistic introductory example
might lead to believe. Indeed, considering Eq. (\ref{eq:criado})
that conserves ``energy'' $e\equiv x^2$, i.e. ${\cal C}(x)=x^2$,
it appears that $p_{\!_E}^{st}(e)
\propto \exp(-\beta e)/\sqrt{e}$ in the steady state \cite{Criado2},
where $\beta$ is some inverse temperature. 
A naive application of Eq. (\ref{eq:SShannon}),
on the other hand, leads to the 
incorrect result 
$p_{\!_E}^{st}(e) \propto \exp(-\beta e)$.
This means that here,
$\Lambda_E(e) \propto \sqrt{e}$, and we learn on this
simple example that the conservation law is not sufficient, 
in general, to obtain the relevant $\Lambda$. This key quantity
is encoded in the transformation law $(x_i,x_j) \to (x'_i,x'_j)$,
in a way that we now bring to the fore. 

To get an idea of the connection ($\Lambda \leftrightarrow $ dynamics),
we first restrict to the subclass of processes that fulfill 
detailed balance. The corresponding single-particle distribution obeys then
\be
\px^{st}(x_1) \px^{st}(x_2) \,dx_1dx_2 
= \px^{st}(x'_1) \px^{st}(x'_2) \,dx'_1dx'_2,
\label{eq:db}
\ee
where, due to the 
mean-field-like sampling procedure with randomly chosen pairs, the two-particle probability distribution factorizes 
for large $N$ into a product of single particle distributions (a more technical
proof will be outlined below). 
On the other hand, assuming that ${\cal Q}$ can be answered 
positively, the  stationary single 
particle distribution $\px^{st}$ 
is constrained to be of the form (\ref{eq:pstar}).
Then, from eqs. (\ref{eq:pstar}),
(\ref{eq:db})  and the conservation law, eq. (\ref{eq:cons}), 
we find that the Jacobian $\cal J$ of the 
transformation $(x_1,x_2)$ to $(x'_1,x'_2)$, admits a factorized form 
\be
{\cal J}(x_1,x_2) \equiv 
\Bigl|\hbox{det}\frac{\partial(x_1',x_2')}{\partial(x_1,x_2)}
\Bigr|
= \frac{\lx(x'_1) \lx(x'_2)}{\lx(x_1) \lx(x_2)}. 
\label{eq:jac}
\ee
We emphasize here that the Jacobian is defined for a given 
value of the stochasticity parameter $\eta$: $x'_1$ and $x'_2$ are
functions of $x_1$, $x_2$, and $\eta$.

We now arrive at our main part and we will show below that if the 
factorization 
property  (\ref{eq:jac}) of $\cal J$ holds (without any other restriction as for example 
detailed balance) then 
the stationary distribution function $\px^{st}$ is of the form (\ref{eq:pstar}),
and hence we are able to answer affirmatively to question $\cal Q$.
In addition, the relevant weighting function $\lx$ can then 
be directly read from (\ref{eq:jac}). This is interesting from an
operational point of view, since the Jacobian directly follows from
the knowledge of the collision law, which is an input of the
model. As a illustration, we return to the toy model of Eq. (\ref{eq:criado}),
recast in the conserved variable $e\equiv x^2$. 
We have ${\cal J}(e_1,e_2) = [e_1'e_2'/(e_1 e_2)]^{1/2}$, which is of the
form (\ref{eq:jac}), with $\Lambda_{_E}(e) \propto \sqrt{e}$.
This immediately leads to the correct distribution $p_{\!_E}^{st}(e) 
\propto \exp(-\beta e)/\sqrt{e}$.

We now proceed with our general proof, that starts with 
assuming property (\ref{eq:jac}) for $\cal J$, and that involves
the following three steps.

a) We introduce a new set of variables,
under the mild assumption that $\lx$ in (\ref{eq:jac})
is non vanishing. Indeed, with the function of $x$ $z(x)=\int^x dx'/\lx(x')$,
the Jacobian of the collision law becomes unity
($dz'_1dz'_2=dz_1dz_2$), which simplifies the kinetic
theory description. 

b) Although our aim is to derive the stationary 
single particle distribution function $\pz$, working at $N$-body level
with the phase space density $\rho_N(\Gamma,t)$, where 
$\Gamma \equiv (z_1,\ldots, z_N)$, turns out to be a convenient
detour. This distribution obeys the following evolution equation 
\cite{rque25,Resibois}
\be
\partial_t \rho_N(\Gamma,t) \,=\, \sum_{i<j}^N \int d\eta w(\eta) 
\left[\,b_{ij}^{(\eta)}-1\right]\rho_N(\Gamma,t)
\label{eq:Nbody}
\ee
where the random variable $\eta$ with distribution $w$ enters
the collision law (see above), that can be described by the 
inverse collision operator $b_{ij}^{(\eta)}$. This operator
acts on the distribution on its right by replacing the arguments
$z_i$ and $z_j$ by their precollisional values $z_i^*$ and $z_j^*$:
\be
b_{12}^{(\eta)}\rho_N(\Gamma,t) = 
\rho_N(z_1^*,z_2^*,z_3,\ldots,z_N,t)
\ee
with $(z_{i,j}^*)'=z_{i,j}$. The present description in terms of $z$ quantities
is also endowed with a conservation law, that we write here --modulo a slight
abuse of notation-- with the same function $\cal C$ as in Eq.
(\ref{eq:cons}): $\sum_i {\cal C}(z_i) = C$.
It is then straightforward to see that the distribution 
$\rho_N^{st} \propto \delta(C-\sum_i {\cal C}(z_i))$ (with proper 
normalization) provides a stationary
solution to Eq. (\ref{eq:Nbody}). 
The corresponding single particle distribution function follows from 
computing the first marginal 
$\px^{st}(z_1) \propto \int \rho_N^{st} dz_2\ldots dz_N$.
The argument is akin to that put forward to construct the canonical 
ensemble from the micro-canonical distribution \cite{Ma}, and leads to
$\pz^{st}(z) \propto \exp[-\beta {\cal C}(z)]$.

c) The last important step in the argument is to show that the $N$-body
measure $\rho_N^{st}$ is attractive, at long times, for arbitrary initial 
conditions sharing the same value of $C$. 
For this purpose, we borrow a technique introduced in \cite{Kac} and consider 
an arbitrary strictly convex positive function $h(x)$ from which we 
construct
\be
H(t) = \int\,d\Gamma \rho_N^{st}(\Gamma) \, h[\rho_N(\Gamma,t)]. 
\ee
The evolution equation (\ref{eq:Nbody}) implies
\begin{eqnarray}\label{eq:dH}
\frac{dH}{dt}= 
\frac{N(N-1)}{2} 
\int d\Gamma d\eta \rho_N^{st}(\Gamma) w(\eta)\nonumber\\
\left\{ h'[\rho_N(\Gamma,t)]
\left[\rho_N(b_{12}^{(\eta)}\Gamma,t)-\rho_N(\Gamma,t)\right]\right.\nonumber\\
\left.+ h[\rho_N(\Gamma,t)]-h[\rho_N(b_{12}^{(\eta)}\Gamma,t)]
\right\}
\end{eqnarray}
where we have used  the invariance under permutation of particle indices and 
that
\begin{equation}
\int d\Gamma d\eta \rho_N^{st}(\Gamma)w(\eta)
\left\{ h[\rho_N(\Gamma,t)]-h[\rho_N(b_{12}^{(\eta)}\Gamma,t)]\right\}=0.
\end{equation}
From the convexity of $h$, we have that $H$ is a non-increasing 
function of time.
Since $H$ is in addition bounded from below, we have established
that it converges at long times to a constant \cite{rque26}. 
Moreover, as the curly bracket of 
eq. (\ref{eq:dH}) only vanishes when 
$\rho_N(\Gamma,t)=\rho_N(b_{12}^{(\eta)}\Gamma,t)$, 
we conclude from our ergodicity assumption that all initial phase space
densities for which the conserved quantity is strictly equal to $C$ evolve
towards $\rho_N^{st}$. This is a flat and finite measure on the ensemble
defined by $\sum_i {\cal C}(z_i)=C$,
and can be
seen as a generalized micro-canonical density. A similar property 
also applies to the first marginal $\pz(z,t)$ that is then
attracted to $\pz^{st} \propto \exp[-\beta {\cal C}(z)]$. 
Returning to the original
variable $x$ in which the problem was formulated, 
and bearing in mind that $dz/dx=\lx^{-1}(x)$, this yields the
desired result that 
the stationary distribution $\px^{st}(x)$ is of the form
(\ref{eq:pstar}). Incidentally, we also obtain here 
that the 2-body distribution $p_{2,X}^{st}(x_1,x_2)$ factorizes in the 
steady state, in the product $\px^{st}(x_1) \px^{st}(x_2)$, 
as mentioned above. 

So far, we have shown that under the assumption (\ref{eq:jac}),
the steady-state distribution $\px^{st}$ can be found by minimizing
a functional of the form (\ref{eq:S}), with a known weighting function
$\Lambda_X$, directly read from (\ref{eq:jac}). This was illustrated
by the toy dynamics (\ref{eq:criado}), but our introductory 
example also may be understood in that framework: with the dynamics
defined by $(x_i,x_j)\to (x_i+\eta,x_j-\eta)$, for which 
${\cal C}(x)=x$, we simply have 
${\cal J}(x_1,x_2) = 1$, hence $\lx=1$ and 
$\px^{st}(x)\propto \exp(-\alpha x)$.

To complete the analysis, three remarks are in order. First, while 
we have restricted to the scalar case for the sake of simplicity,
$x$ can equally be a vectorial quantity.
Second, 
Eq. (\ref{eq:jac}), when it applies, does not define a unique
function $\lx$. Indeed, consider two candidates obeying
\begin{equation}
\frac{\lx(x'_1) \lx(x'_2)}{\lx(x_1) \lx(x_2)} \, =\, 
\frac{\widetilde{\Lambda}_{\!_X\!}(x'_1)\widetilde{\Lambda}_{\!_X\!}(x'_2)}
{\widetilde{\Lambda}_{\!_X\!}(x_1)\widetilde{\Lambda}_{\!_X\!}(x_2)}
\end{equation}
for all $x_1$, $x_2$, and $\eta$. Then, $\log(\lx/\widetilde{\Lambda}_{\!_X\!})$
is a collisional invariant. Assuming that there is no
``hidden'' conservation law, we have
$\log(\lx(x)/\widetilde{\Lambda}_{\!_X\!}(x)) = a + b \,{\cal C}(x)$,
where $a$ and $b$ are arbitrary constants.
So, a candidate weighting function defined through (\ref{eq:jac}),
is prescribed up to a function $\exp(a+ b\,{\cal C}(x))$.
Such a freedom in the choice of $\Lambda$ only shifts 
the functional (\ref{eq:S}) by the constant $a+b\langle {\cal C}\rangle$. 
The final result for $\px^{st}$ is hence not affected
by the choice made for $\lx$.

Third, it seems worthwhile to provide 
a more intuitive understanding of the fact that if the Jacobian 
$(x_1,x_2)\to(x'_1,x'_2)$ fulfills Eq. (\ref{eq:jac}), 
then the corresponding
weighting function is proportional to the $\lx$
appearing in (\ref{eq:jac}). 
For any pair $(x_1,x_2)$, (\ref{eq:jac}) implies that
the respective uncertainties
$\delta x_1$ and $\delta x_2$, will be affected by the collision 
such that $\lx(x_1)^{-1}\delta x_1 \lx(x_2)^{-1}\delta x_2 
= \lx(x_1')^{-1}\delta x_1' \lx(x'_2)^{-1}\delta x'_2$.
At long times, the system reaches a state where the solution 
to the above constraint is simply $\delta x/\lx(x)=\hbox{cst}$,
so that at one body level, 
the space of dynamical variables is resolved with an $x$-dependent precision 
$\delta x \propto \lx(x) $. This allows to view $\lx$ as
an $x$-dependent volume in the space of dynamical variable,
that quantifies the ``graining'' with which the
space is resolved. 
Alternatively, this argument shows
that the density of points generated in $x$-space verifies
$\mx(x) \propto 1/\delta x \propto 1/\lx(x)$,
as already mentioned.

To summarize, we have studied a class of problems 
encountered in different contexts, 
such as soft matter where a mixture 
of polydisperse hard spheres \cite{Z99,B00,BC01}, hard rods \cite{EMPT10},
or ring polymers \cite{CS02}
have been shown to exhibit a condensation in real space,
stochastic mass transport models \cite{M09}, or in mathematical
literature where the Kac walk \cite{Kac} is an important
kinetic theory toy model for studying the propagation of chaos and rate of
equilibration \cite{C10}.
Specifically, our goal here was to analyze under which conditions the steady 
state distribution $\px^{st}$
obtained by iterating a generic collision process with
conservation law [Eq. (\ref{eq:cons})] could
equivalently be obtained from a maximum entropy argument by extremalizing a
given functional of the type (\ref{eq:S}). 
We have found 
that this is the case 
if the Jacobian 
of the collision law $(x_1,x_2)\to(x'_1,x'_2)$, 
can be written as in (\ref{eq:jac}), from which the relevant 
weighting function $\lx(x)$ can be extracted,
which provides simply $\px^{st}$. This is for example 
the case of Refs. \cite{Z99,CS02,D01,Criado,B00,BC01,EMPT10}. 
The connection
thereby established is free of the so-called measure problem,
that plagues a naive writing of the entropy functional as in 
(\ref{eq:SShannon}), an expression first proposed by Shannon,
and that propagated in a significant fraction of the literature.
Our analysis, in other words, provides the correct prior
$\lx^{-1}$ that should be considered, see Eq. (\ref{eq:S}).  
A key point is that the configurations allowed by the conservation law(s)
are in general sampled non uniformly. This non-uniformity,
encoded in the $x$ dependence of $\lx$,
that gives different weights to different points in $x$-space,
is the feature ensuring that the information measure considered is
absolute, and does not depend on the parameterization chosen. 
We finally note that our approach --which includes
multiple conservation laws-- can be generalized to more
complex collisional processes, involving more than two bodies,
or in which the collision frequency $\omega(x_1,x_2)$,
chosen constant here for the sake of simplicity, 
actually depends on the pair considered, as long as
$\omega(x_1,x_2) = \omega(x'_1,x'_2)$.




\begin{thebibliography}{10}
\bibitem{rquesha}
When quoting Shannon's seminal work \cite{Shannon}, we do not refer to
his well known derivation of the entropy or information content of a discrete
probability set, but to his generalization
to continuous distributions (part III section 20 of Ref \cite{Shannon}), 
that suffers from the measure problem discussed here. 
In addition, the example discussed here with \smash{$\px=e^{-x}$}
can be found in section 20.7 of that paper.
\bibitem{rque2}
We note that Eq. (\ref{eq:S}) is reminiscent of a
Kullback-Leibler distance (also called relative entropy) \cite{CT}, with the 
difference that $\mx \propto 1/\lx$ is not a probability density.
\bibitem{Jaynes2}
See ref \cite{Jaynes}, page 203.
There is consequently a rather large body of literature that 
erroneously considers the
measure problem --absent in the discrete case-- as 
intrinsic to the continuous formulation, leaving aside the textbooks
that do not mention the issue.
\bibitem{rque25}
Here, time $t$ counts the number of collisions occurred.
The problem can be recast in a continuous $t$ form.
\bibitem{rque26}
Note that the fact that $H$ is non-increasing does not provide
an $H$-theorem in the usual sense \cite{Ma,Reichl}, since 
our function depends on the final distribution. Additionaly,
because of the lack of detailed balance (which translates
into $z_i^* \neq z_i'$), it is not possible to obtain directly an
$H$ functional at the single particle level.
\end{thebibliography}
\end{document}